# Work Function of Bismuth Telluride: First-Principles Approach

**Byungki Ryu**[*]

*Thermoelectric Conversion Research Center, Korea Electrotechnology Research Institute (KERI),*

*Changwon 51543, Republic of Korea*

First-principles approach is demonstrated to calculate the work function of $Bi_2Te_3$. The reference potential and the vacuum energy levels are extracted from the $Bi_2Te_3$ (0001) surface structure using the reference potential method based on the density functional theory. The one-shot $G_OW_O$ many-body perturbation theory is used to place the bulk band edge energies with respect to the reference level and the vacuum energy. At last, the work function of 5.301 – 5.131 eV is predicted for $Bi_2Te_3$ (0001) surface and compared to various elements.

## I. INTRODUCTION

$Bi_2Te_3$-related binaries and ternaries have attracted much attention due to its use for thermoelectric materials[1,2,3] as well as its unique band structure exhibiting topological insulating surface state.[4,5] The strong spin-orbit-interaction (SOI) from heavy element nature strongly affects its band structure to have non-trivial band structure as well as multiple band valley degeneracy.[4,6,7] With the importance of this material, there have been a lots of studies to investigate the bulk and surface band structures of $Bi_2Te_3$.[4,7-14]

In thermoelectrics, the complex band structure is very critical for high thermoelectric energy conversion efficiency due to the large anisotropic effective mass.[1,2,15] The band structure calculations based on the density-functional-theory (DFT)[16,17] well describe this band structure anisotropy in $Bi_2Te_3$.[12,18,19] However, due to the band gap problem in DFT originating from derivative discontinuity in exchange-correlation energy,[20] the band gap ($E_g$) of $Bi_2Te_3$ is also underestimated to be less than ~100 meV.[7,9,10] Recent studies revealed that the band structure as well as the band gap are sensitive to the lattice constant, structure relaxation, and the selection of the exchange correlation energy.[7-12] As a result, the thermoelectric transport properties form electron is severely underestimated for high temperature or intrinsic doping region.[21] To overcome the band gap problem, the self-energy is corrected by using the hybrid-DFT[8,22] and the many-body-perturbation-theory (MBPT) based GW calculations.[12-14] As a result, the quasi particle band gap is doubled compared to DFT results, consistent to the experimental band gap (0.17 eV).[12]

The band alignment is one of the important physical properties in composite materials or in the devices.[23-28] In solar-cell, the band alignment in device affects the chare separation to enhance the energy conversion efficiency.[29] In thermoelectric material, the band alignment in thermoelectric composite materials is very important, affecting the carrier transport mechanism.[30-34] For example, the nano-sized metal in thermoelectric material acts as a potential barrier to change the electron relaxation time.[32] The energy dependent transmission through metal nanoparticle can act as the energy filter for charge transport and the power factor can be enhanced especially in super-lattice structure.[32,34] Recent study also reported the possibility of minority carrier blocking in hetero nanowire thermoelectric material, enhancing high temperature thermoelectric property.[35-38]

On one hand, the band gap problem in DFT affects the absolute position of band edge levels with respect to the vacuum level, resulting in the wrong band alignments and work functions in the interfaces or surfaces. In Si/SiO$_2$ interface structure, the valence band offset is larger than ~4eV.[39,40] However, in DFT, they are underestimated by ~1eV in DFT.[41] In the case of Si/HfO$_2$ interface structure, the conduction band offset is severely underestimated and become negligible, even if the HfO$_2$ insulation layer well acts as potential barrier for electron carrier in Si-based devices.[42,43] We speculate that the wrong band alignment may lead to the wrong design of thermoelectric materials and composites.

Even though the band structure and band alignments of materials are important, only the band gap and the effective mass of materials were theoretically studied for Bi$_2$Te$_3$. There is lack of studies on work function or band alignments for Bi$_2$Te$_3$. In 1959, the work function of Bi$_2$Te$_3$ is reported to be ~5.3 eV under p-type condition.[44] However, in our knowledge, there is no other study for Bi$_2$Te$_3$ work function measurement as well as theoretical prediction. Here, by performing the density functional theory and the quasi particle GW calculations, we successfully calculate the work function of the Bi$_2$Te$_3$ (0001) surface. The calculated work function is consistent to the value of work function measurement for Bi$_2$Te$_3$ (0001) cleavage plane.[44] At last, based on the work function of Bi$_2$Te$_3$, we classify the elements as p- and n-metal for the Ohmic interface between Bi$_2$Te$_3$ and elements for the future studies

## II. CALCULATION METHOD

For DFT, we use the generalized-gradient-approximation parameterized by Perdew, Burke, and Ernzerhof (PBE) [45] for the exchange-correlation potential, projector-augmented-wave pseudopotential,[46] implemented in VASP code.[47,48] For G$_O$W$_O$ approximation,[49,50] we use the one-particle Hamiltonian from PBE to calculate the green function G and screened Coulomb potential W without any update. Then the self-energy is calculated from G and W and the quasi-particle band structure is calculated. Note that, here we neglect the off-diagonal term in the self-energy. In all DFT and GW calculations, we include the spin-orbit-interaction (SOI).

The work function of $Bi_2Te_3$ (0001) surface is calculated using the reference potential method.[51] The DFT calculation is used to calculate the reference potential of $Bi_2Te_3$ ($E_{ref}^{surf}$) and the vacuum energy ($E_{vac}^{surf}$). The MBPT-based one-shot GW ($G_OW_O$) calculation is used to calculate the position of the mid-gap energy ($E_m^{bulk}$) as well as band edge energies with respect to the reference potential for $Bi_2Te_3$ ($E_{ref}^{bulk}$). Thus, two separate calculations are performed: one is for the bulk property with $G_OW_O$ band edge correction and the other is for the surface property with DFT electrostatic potential. Then, the mid-gap energy is calculated with respect to the vacuum energy level. Then, the work function ($E_{WF}$) can be written as $E_{WF} = E_{vac} - E_m + \mu = (E_m - E_{ref})^{bulk} - (E_{vac} - E_{ref})^{surf} + E_F$, where $\mu$ is the electron chemical potential ranging from $-E_{gap}/2$ for p-type to $E_{gap}/2$ for n-type. For the band gap, we use the experimental value of 0.17 eV.[12]

For the model structure of bulk and surface $Bi_2Te_3$, we use the experimental lattice parameters for $Bi_2Te_3$ structure ($a_{hex}^{bulk}$ = 4.3835 Å, $c_{hex}^{bulk}$ = 30.487 Å). [52] The internal atomic coordinates are fully relaxed for bulk structure with an energy cutoff of 175 eV and a **k**-point mesh of 12×12×12. The obtained internal parameters are $u$ = 0.4005 and $v$ = 0.2097. Previous studies revealed that the k-point sampling is very critical to describe the band structure and the value of the band gap. Thus, for DFT calculations, we use the sufficiently dense k-point mesh of 12×12×12 for bulk $Bi_2Te_3$. However, due to the large computational cost of MBPT-based GW calculation, we use the coarse **k**-point mesh of 6×6×6 for bulk calculations. Here, all **k**-point meshes are uniform and Γ-centered.

To model the $Bi_2Te_3$ (0001) surface structure, we use the supercell approach. In the supercell, 1 to 6 quintuple layers (QLs) of $Bi_2Te_3$ are contained with a sufficient vacuum region larger than 3 nm. The used lattice parameters for supercell are $a_{hex}^{surf-super}$ = 4.3835 Å, $c_{hex}^{surf-super}$ = ($c_{hex}^{bulk}$/3) × 9 = 91.461 Å. Note that we neglect the relaxation along *c*-direction, considering the weak interaction between the adjacent $Bi_2Te_3$ QLs and negligible distortion at the surface.[53] The reference potential and the vacuum energy levels are calculated with the energy cutoff of 400 eV and the **k**-point mesh of 12×12×1.

## III. RESULTS AND DISCUSSION

First, we investigate the bulk band structure of $Bi_2Te_3$. The calculated band energies for VBM and CBM ($E_{VBM}$ and $E_{CBM}$), the mid gap energy $E_m$, and the band gap $E_g$ are shown in **Table 1**. The calculated band gaps are sensitive to the **k**-point mesh and the inclusion of SOI for PBE and $G_0W_0$ calculations, consistent to the previous work.[7] Note that the $E_g$s are also sensitive to the number of bands (NB) for $G_0W_0$ calculations, as discussed in other works.[54] In PBE plus SOI calculations (denoted as PBE+SOI), the band gap is calculated to be 0.185 eV with coarse **k**-point mesh of (6×6×6) and it is 0.105 eV with fine **k**-point mesh of (12×12×12). We would like to note that, in contrast to the band gap, the mid gap energy is less sensitive to the **k**-point sampling. The difference of $E_m$ between the coarse and fine meshes is only 12 meV. When we correct the band edges and calculate the work function, we use the $E_m$ instead of band edge energies because of the huge computational cost in $G_0W_0$ calculations with fine **k**-point mesh. Meanwhile, the band gap is less sensitive to the cutoff energy. When $E_{cut}$ is increased from 175 eV to 400 eV, the band gap is slightly decreased by 4 meV.

In MBPT-based GW calculations, the number of empty bands is very important to determine the band gap and band edge energies.[54] So we perform the convergence test for various NB values with the coarse **k**-point mesh of (6×6×6). Fortunately, the $E_g$ is rapidly converges within 10 meV. The $G_0W_0$ with initial Hamiltonian from PBE plus SOI (denoted as $G_0W_0$+SOI) $E_g$ is 0.249 eV when NB = 240 and it is rapidly converged to 0.257 eV when NB = 1200. In contrast, even we use very large NB of 1200, which is 64.3 times the number of occupied bands (28), the band edges are not converged yet. As shown in the **Figure 1**, The band edge and the mid-gap energies are linearly extrapolated with the equation of $Y = C_1X + C_2$, where X is 1/NB, Y is band energy, $C_n$ are the fitted coefficient. At last, we obtain the $E_m$ of 3.761 eV. We also perform the fine **k**-point mesh calculation and obtain the the $G_0W_0$ band gap of 0.124 eV with NB = 120, however, the calculated band gap is still far from the experimental band gap (0.17 eV). Due to the computational cost, we can not test the NB convergence for fine-mesh calculations. For the band gap correction, we use the experimental band gap rather use the $G_OW_O$ band gap. Thus, the band gap corrected VBM and CBM energies are estimated as $E_m - 1/2\ E_g$ and $E_m + 1/2\ E_g$, respectively.

We would like to mention the computational time ($t_{comp}$) of one-shot $G_OW_O$ calculations (see the last coulomb in **Table 1**). Here $t_{comp}$ is measured as one-node computation time in KERI's High-Performance-Computing (HPC) server.[55] The $t_{comp}$ is only or less than a few minutes for PBE calculations including dielectric function calculations with lots of unoccupied bands. However, $t_{comp}$ is extremely large for $G_OW_O$. Note that the $t_{comp}$ is exponentially increased when **k**-point mesh changes. The computational time of $G_OW_O$+SOI with NB = 120 for (6×6×6) **k**-point mesh is 6.3 hours and it is tremendously increased to 90 days (2167 hours) for (12×12×12) **k**-point mesh. Thus, due to the huge computational cost, we only consider the coarse **k**-point mesh for band structure from $G_OW_O$ calculation. Also see that the inclusion of SOI also doubles the computational cost.

Next, we investigate the band structure of $Bi_2Te_3$ (0001) surface structure with various QL numbers (*N*s). In **Figure 2**, we show the band energies, $E_m$, and Fermi level ($E_F$) for *N*-QL $Bi_2Te_3$ structure, where *N* changes from 1 to 6, within PBE calculations. When $N = 1$, the band gap is 0.3 eV. When $N = 2$, the band gap is reduced to ~0.1 eV. When $N = 3$, there is a negative band gap due to the band inversion. When N > 3, the band gap is totally inverted and the topological surface states are formed within the band gap. Note that, for N > 3, the surface and bulk states can be decoupled. Here, VBM and CBM denote the bulk band edge states, not surface states. We also check the positions of Dirac-cones (DCs) in $Bi_2Te_3$ surface states. Here, due to the low position of the Dirac-cone (DC) below the VBM level and the half-filled surface states, the $E_F$ is located below the mid-gap position. Please, note that the $E_m$ and $E_F$ are converging rapidly, while $E_{CBM}$, $E_{VBM}$, and $E_g$ are converging slowly when *N* goes from 3 to 6. Considering the fast convergence of $E_m$ and $E_F$, we use the 6-QL surface structure for the work function calculations.

Then, we calculate the reference level from the average local potential in $Bi_2Te_3$ (0001) surface structure in PBE+SOC level. From the above results, we see that when there are 6-QLs in the surface structure the $E_m$ and $E_F$ are converged sufficiently. In **Figure 3(a) and (b)**, we show the ball-and-stick model for the 6-QL $Bi_2Te_3$ (0001) surface structure and draw the average-local-potential along $z$-direction ($c$-direction), defined as $V(z) = ( \int d\mathbf{x} d\mathbf{y}\ V(\mathbf{x},\mathbf{y},\mathbf{z}) ) / ( \int d\mathbf{x} d\mathbf{y} )$. Here the ball-and-stick atomic structure is visualized by the VESTA program.[53] The reference potential of $Bi_2Te_3$ is then calculated by averaging the $V(z)$ over the centered-two-QL region, defined as $E_{ref} = ( \int_{\text{center-two-QL}} V(\mathbf{z})\ d\mathbf{z} ) / ( \int_{\text{center-two-QL}} d\mathbf{z} )$. The $E_{ref}$ is positioned at 8.965 eV below the vacuum energy level.

Finally, we calculate the work function of $Bi_2Te_3$ (0001) surface by aligning the reference level in bulk and surface. From $G_OW_O$+SOI, the energy difference $E_1$ between the $E_m$ and $E_{ref}$ are extrapolated to be 3.761 eV. Since the energy difference $E_2$ between $E_{vac}$ and $E_{ref}$ is 8.965 eV, the mid-gap energy with is located at 5.204 eV below the vacuum level ($E_{WF} = E_2 - E_1 = E_{vac} - E_m$). Considering the band gap of $Bi_2Te_3$ (0.17 eV), we calculate the work function to be 5.204 + 0.17 / 2 = 5.289 eV for p-type and 5.119 eV for n-type. To see the effect of SOI and MBPT mid-gap energy correction, we summarize the mid-gap work function results from PBE, PBE+SOI, $G_OW_O$, and $G_OW_O$+SOI results in **Table 2**. See that, without $G_OW_O$, the work functions are under estimated from experimental value by ~0.3 eV. We also correct the k-point mesh problem in GW calculations. Our $G_OW_O$ calculations use the coarse k-point mesh. To correct the k-point mesh effect, we add the correction estimated from PBE+SOC calculations. In PBE+SOC, the work function difference between coarse and fine **k**-point meshes is 12 meV. Thus, we add this difference to correct the coasre **k**-point mesh effect in $G_OW_O$+SOI and obtain the 5.301 eV for p-type WF, consistent to the experimental value of 5.3 eV.

The work function difference between two materials is very important because it can be related to the interfacial properties such as the type of junction (Ohmic contact or Schottky contact), the height of Schottky barrier, and charge transfer between materials. It is also very important in thermoelectric composite composed of thermoelectric material with metal nano-inclusion due to the impact on the electron scattering at the interface. When the metal nanoparticles were incorporated in the thermoelectric materials, the charge carriers can be scattered at the semiconductor-metal interface due to the barrier height or due to the band bending. Here, by comparing the calculated work function of $Bi_2Te_3$ and 63 elements, we classify the elements as the p-and n-metals for $Bi_2Te_3$-metal contacts (see **Figure 4**). The WF values are obtained from the website Wikipedia.[54] If the material's work function is larger than or equal to the WF of p-$Bi_2Te_3$, it is the p-metal material. If the material's work function is smaller than or equal tot WF of n-$Bi_2Te_3$, it is the n-metal. Note that there are small number of p-metal elements (Se, Pt, Pd, Ir, and Au), compared to the number of n-metal elements. Also note that Ni might cause the Fermi-level pinning at the $Bi_2Te_3$/Ni interface

## IV. CONCLUSION

In conclusion, first-principles calculations are demonstrated to calculate the work function of $Bi_2Te_3$ (0001) surface. The calculated work function is very sensitive to the choice of computational method. By using the one-shot $G_OW_O$ calculations, we finally obtain the p-type work function of 5.301 eV and n-type work function of 5.216 eV, consistent to the experimentally measured one. Also we classify the elements as the p- and n-metals for $Bi_2Te_3$-metal contacts.

## ACKNOWLEDGEMENT

This research was supported by the Development of Middle-high Temperature Thermoelectric module for Nuclear Battery project by Ministry of Science, ICT & Future Planning (2017M1A3A9015334).

## REREFENCES

[1] H. J. Goldsmid, Introduction to Thermoelectricity (Springer, Berlin, Heidelberg, 2010)

[2] D. M. Rowe, Thermoelectrics Handbook (CRC Press, Boca Raton, 2006).

[3] S. I. Kim, K. H. Lee, H. A. Mun, H. S. Kim, S. W. Hwang, J. W. Roh, D. J. Yang, W. H. Shin, X. S. Li, Y. H. Lee, G. J. Snyder, and S. W. Kim, Science **348**, 109 (2015).

[4] H. Zhang, C.-X. Liu, X.-L. Qi, X. Dai, Z. Fang, and S.-C. Zhang, Nature Phys. **5**, 438 (2009).

[5] Y. L. Chen, J. G. Analytis, J.-H. Chu, Z. K. Liu, S.-K. Mo, X. L. Qi, H. J. Zhang, D. H. Lu, X. Dai, Z. Fang, S. C. Zhang, I. R. Fisher, Z. Hussain, and Z.-X. Shen, Science **325**, 178 (2009).

[6] H. Shi, D. Parker, M.-H. Du, and D. J. Singh, Phys. Rev. Applied **3**, 014004 (2015).

[7] B. Ryu, M.-W. Oh, B.-S. Kim, J. E. Lee, S.-J. Joo, B.-K. Min, H. W. Lee, and S. D. Park, J. Kor. Phys. Soc. **68**, 115 (2016).

[8] S. Park and B. Ryu, J. Kor. Phys. Soc. **69**, 1683 (2016).

[9] S. J. Youn and A. J. Freeman, Phys. Rev. B **63**, 085112 (2001).

[10] G. Wang and T. Cagin, Phys. Rev. B **76**, 075201 (2007).

[11] X. Luo, M. B. Sullivan, and S. Y. Quek, Phys. Rev. B **86**, 184111 (2012).

[12] E. Kioupakis, M. L. Tiago, and S. G. Louie, Phys. Rev. B **82**, 245203 (2010).

[13] I. A. Nechaev and E. V. Chulkov, Phys. Rev. B **88**, 165135 (2013).

[14] O. V. Yazyev, E. Kioupakis, J. E. Moore, and S. G. Louie, Phys. Rev. B **85**, 161101(R) (2012).

[15] H. J. Goldsmid, Proc. Phys. Soc. 71, 633 (1958).

[16] P. Hohenberg, and W. Kohn, Phys. Rev. **136**, B864 (1964).

[17] W. Kohn and L. J. Sham, Phys. Rev. 140, A1133 (1965).

[18] P. Larson, S. D. Mahanti, M. G. Kanatzidis, Phys. Rev. B **61**, 8162 (2000).

[19] B. Ryu and M.-W. Oh, J. Kor. Cer. Soc. 53, 273 (2016).

[20] J. P. Perdew, Int. J. Quantum Chem. 28, 497 (1985).

[21] B. Ryu, J. Chung, E.-A. Choi, B.-S. Kim, and S.-D. Park, J. Alloy. Comp. 727, 1067 (2017).

[22] M. Kim, A. J. Freeman, C. B. Geller, Phys. Rev. B **72**, 035205 (2005).

[23] D.-H. Choe, D. West, S. Zhang, arXiv:1705.04432 (2017).

[24] H. Kroemer, Rev. Mod. Phys. 73, 783 (2001).

[25] S. M. Sze and K. K. Ng, Physics of semiconductor devices (Wiley, Hoboken, NJ, ed. 3, 2007).

[26] V. S. Bagotsky, Ed., Fundamentals of electrochemistry (Wiley, Hoboken, NJ, ed. 2, 2005).

[27] A. J. Bard and L. R. Faulkner, Electrochemical methods: fundamentals and applications (Wiley, New York, NY, ed. 2, 2001).

[28] R. T. Tung, The physics and chemistry of the Schottky barrier height. Appl. Phys. Rev. 1, 011304 (2014).

[29] C-.H. M. Chung, P. R. Brown, V. Bulović, and M. G. Bawendi, Nat. Mater. 13, 796 (2014).

[30] L. D. Zhao, J. He, S. Hao, Chun-I Wu, T. P. Hogan, C. Wolverton, V. P. Dravid, and M. G. Kanatzidis, J. Am Chem. Soc. 134, 16327 (2012).

[31] J. M. O. Zide, D. Vashaee, Z. X. Bian, G. Zeng, J. E. Bowers, A. Shakouri, and A. C. Gossard, Phys. Rev. B **74**, 205335 (2006).

[32] S. V. Faleev and F. Léonard, Phys. Rev. B **77**, 214304 (2008).

[33] S. Hwang, S.-I. Kim, K. Ahn, J. W. Roh, D.-J. Yang, S.-M. Lee, and K.-H. Lee, J. Electron. Mater. 42, 1411 (2012).

[34] M. Thesberg, M. Poufath, N. Neophytou, and H. Kosina, J. Elecron. Mater. 45, 1584 (2016).

[35] J.-H. Bahk and A. Shakouri, Appl. Phys. Lett .105, 052106 (2014).

[36] H. Yang, J.-H. Bahk, T. Day, A. M. S. Mohammed, G. J. Snyder, A. Shakouri, and Y. Wu, Nano Lett. 15, 1349 (2015).

[37] P. G. Burke, B. M. Curtin, J. E. Bowers, and A. C. Gossard, Nano Energy 12, 735 (2015).

[38] J.-H. Bahk and A. Shakouri, Phys. Rev. B 93, 165209 (2016).

[39] J. L. Alay, J. Appl. Phys. 81, 1606 (1997).

[40] R. Shaltaf, G.-M. Rignanese, X. Gonze, F. Giustino, and A. Pasquarello, Phys. Rev. Lett. 100, 186401 (2008).

[41] T. Yamasaki, C. Kaneta, T. Uchiyama, T. Uda, and K. Terakura, Phys. Rev. B 63, 115314 (2001).

[42] E.-A. Choi and K. J. Chang, Appl. Phys. Lett. 94, 122901 (2009).

[43] B. Ryu and K. J. Chang, Appl. Phys. Lett. 97, 242910 (2010).

[44] D. Haneman, J. Phys. Chem. Solids, 11, 205 (1959).

[45] J. P. Perdew, K. Burke, and M. Ernzerhof, Phys. Rev. Lett. **77**, 3865 (1996).

[46] P. E. Blöchl, Phys. Rev. B **50**, 1793 (1994).

[47] G. Kresse and J. Furthmüller, Phys. Rev. B **54**, 11169 (1996).

[48] G. Kresse and J. Joubert, Phys. Rev. **59**, 1758 (1999).

[49] M. S. Hybertsen and S. G. Louie, Phys. Rev. B, 34, 5390 (1986).

[50] M. van Schilfgaarde, T. Kotani, and S. Faleev, Phys. Rev. Lett. 96, 226402 (2006).

[51] C. G. van der Walle and R. M. Martin, Phys. Rev. B 35, 8154 (1987).

[52] W. G. Wyckoff, Crystal Structures (Wiley-Interscience, New York, 1964).

[53] D. Haneman, Phys. Rev. 119, 567 (1960).

[54] B.-C. Shih, Y. Xue, P. Zhang, M. L. Cohen, and S. G. Louie, Phys. Rev. Lett. 105, 146401 (2010).

[55] KERI's HPC server is composed of ~40 nodes. Each computing node includes two Intel Xeon processor (E5-2640v4) with a memory of 128 GB. Each GW calculation used one-computing node and the computational time was measured.

[53] K. Momma and F. Izumi, J. Appl. Crystallogr. 44, 1272 (2011).

[54] Element's work functions are obtained from the website (If the work function is the range, I use the middle value), https://en.wikipedia.org/wiki/Work_function , (Accessed 14 Nov. 2017).

**Table 1** The band edge energies $E_{VBM}$ and $E_{CBM}$, the mid gap energy $E_m$, and the band gap $E_g$ are calculated for the $Bi_2Te_3$ rhombohedral primitive unit cell, using the DFT-PBE and one-shot $G_0W_0$ calculations. The computational cost also calculated in unit of 1 node time in KERI's high-performance-computation (HPC) cluster.[55] The band edge energies are calculated with respect to the reference potential, defined as the average local electrostatic potential over the primitive cell.

|  | k-point mesh | Number of bands | $E_{VBM}$ | $E_{CBM}$ | $E_m$ | $E_g$ | Computational Time |
|---|---|---|---|---|---|---|---|
| PBE | 6×6×6 | 120 | 4.177 | 4.460 | 4.318 | 0.282 | 0.8 minutes |
| PBE | 6×6×6 | 240 | 4.177 | 4.460 | 4.318 | 0.282 | 1.3 minutes |
| $G_0W_0$ | 6×6×6 | 120 | 4.340 | 4.801 | 4.571 | 0.461 | 2.9 hours |
| $G_0W_0$ | 6×6×6 | 240 | 4.103 | 4.593 | 4.348 | 0.490 | 5.2 hours |
| $G_0W_0$ | 6×6×6 | 360 | 4.005 | 4.502 | 4.253 | 0.496 | 7.4 hours |
| $G_0W_0$ | 6×6×6 | 480 | 3.956 | 4.459 | 4.207 | 0.503 | 8 hours |
| PBE+SOI | 6×6×6 | 120 | 4.012 | 4.196 | 4.104 | 0.185 | 1.6 minutes |
| $G_0W_0$+SOI | 6×6×6 | 120 | 4.560 | 4.801 | 4.681 | 0.241 | 6 hours |
| $G_0W_0$+SOI | 6×6×6 | 240 | 4.163 | 4.412 | 4.287 | 0.249 | 10 hours |
| $G_0W_0$+SOI | 6×6×6 | 1200 | 3.737 | 3.993 | 3.865 | 0.257 | 2 days |
| $G_0W_0$+SOI | 6×6×6 | *infinite* | 3.631 | 3.891 | 3.761 | 0.259 | (*extrapolation*) |
| PBE+SOI | 12×12×12 | 120 | 4.039 | 4.144 | 4.092 | 0.105 | 5 minutes |
| $G_0W_0$+SOI | 12×1×12 | 120 | 4.631 | 4.755 | 4.693 | 0.124 | 90 days |

**Table 2** The work function is predicted for various conditions (p-type $Bi_2Te_3$, n-type $Bi_2Te_3$, and intrinsic $Bi_2Te_3$ where Fermi level is at the mid gap) from various calculations In PBE and $G_OW_O$ calculations, the band edge position does not depend the k-point mesh due to the direct band gap at Γ. In PBE+SOI and $G_OW_O$+SOI calculations, the band edge positions are sensitive to the k-point mesh as well as the self-energy calculation method (PBE or GW). The k-point mesh correction is done using the equation f = e+c-a.

|  | Calculated Work Function, $E_{WF}$ | | | |
| --- | --- | --- | --- | --- |
|  | **Mid-gap** $(E_{vac} - E_m)$ | **p-type** $(E_{vac} - E_m + 1/2 E_g)$ | **n-type** $(E_{vac} - E_m - 1/2 E_g)$ | **k-point mesh** |
| **PBE (a)** | 4.648 | 4.733 | 4.563 | Γ-centered |
| **PBE+SOI (b)** | 4.861 | 4.946 | 4.776 | 6×6×6 |
| **PBE+SOI (c)** | 4.873 | 4.958 | 4.788 | 12×12×12 |
| **G_OW_O (d)** | 4.899 | 4.984 | 4.814 | Γ-centered |
| **G_OW_O+SOI (e)** | 5.204 | 5.289 | 5.119 | 6×6×6 |
| **G_OW_O+SOI (f)** | *5.216* | *5.301* | *5.131* | 12×12×12 (**f = e+c-a**) |
| **Expt.** |  | 5.3 |  | **Ref. 44** |

**Figure Captions**

**Figure 1.** The band edge and the mid gap energies of bulk $Bi_2Te_3$ are calculated from the one-shot $G_OW_O$+SOI calculations with coarse k-point mesh (6×6×6). These energies are extrapolated for zero 1/NB (infinite. The average local potential of bulk $Bi_2Te_3$ is set to zero. In PBE+SOC calculations with coarse k-mesh, the VBM and CBM energies are 4.011 and 4.196 eV, respectively.

**Figure 2.** The band edges (VBM an CBM), the mid gap, the Fermi level (EF), and the Dirac-cone (DC) energies of $Bi_2Te_3$ (0001) surface structure are calculated from the PBE+SOI. Here we consider various $N$ QL structures ($N$=1 to 6). When $N$ exceeds 2 or 3, the band gap is inverted and the surface states are formed right below the bulk VBM. Meanwhile, the bulk Fermi-level position is rapidly converged, while the VBM and CBM energy values are changing with $N$.

**Figure 3.** (a) Atomic structure of $Bi_2Te_3$ (0001) surface structure with 6-QLs. Here vacuum size is ~3 nm. (b) The average local potential is calculated and plotted along the z-direction. The vacuum energy is set to zero.

**Figure 4.** Work functions of $Bi_2Te_3$ with various conditions (intrinsic, p-type, and n-type) are drawn with element work functions. P-metal and N-metal elements are in orange (or red) and blue, respectively. Otherwise, it is in green. The values in parenthesis are corresponding work functions.

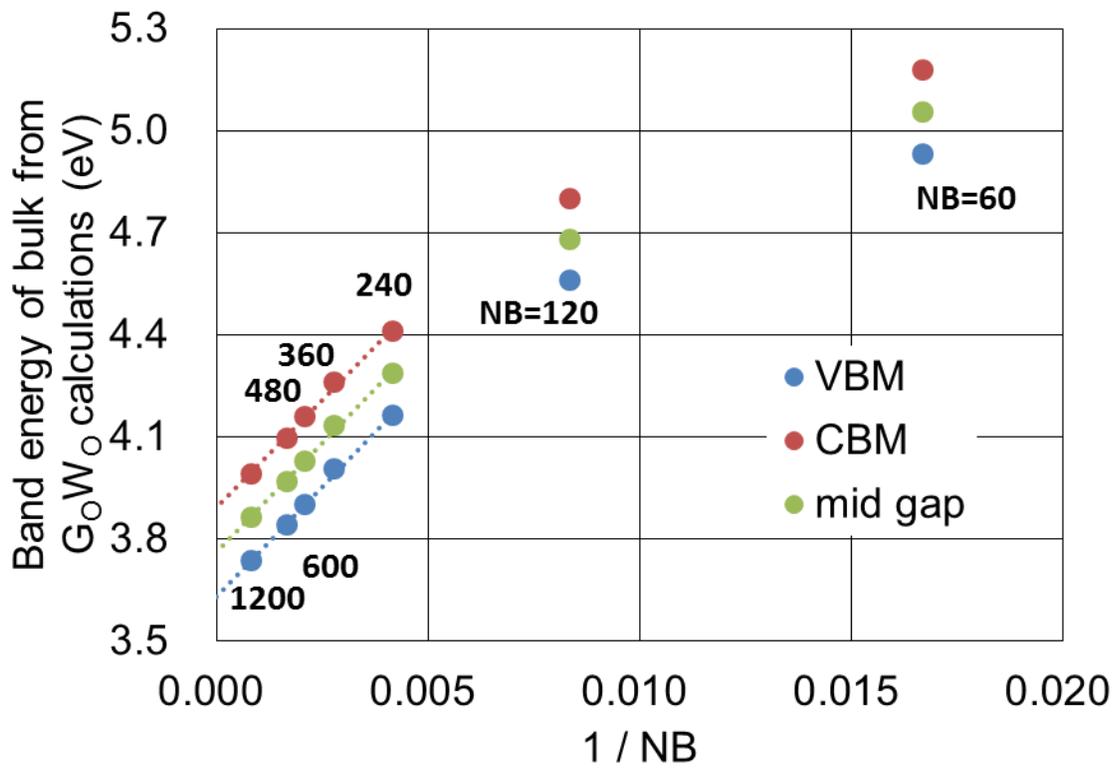

**Figure 1**

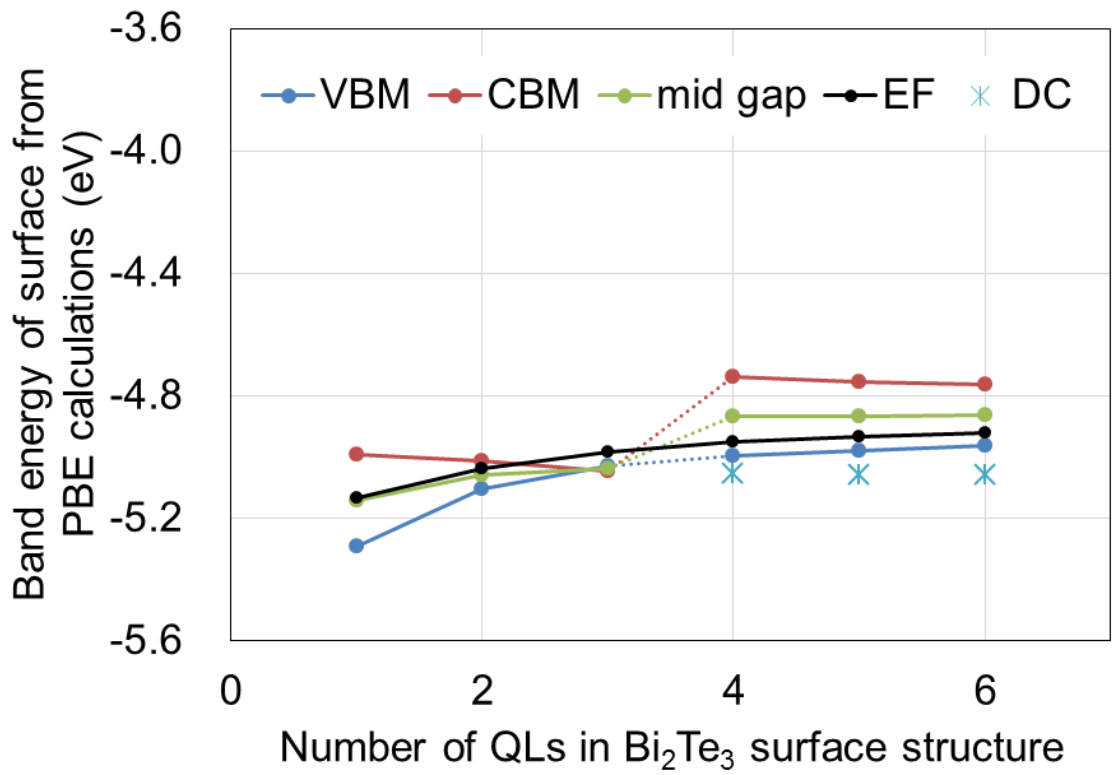

**Figure 2**

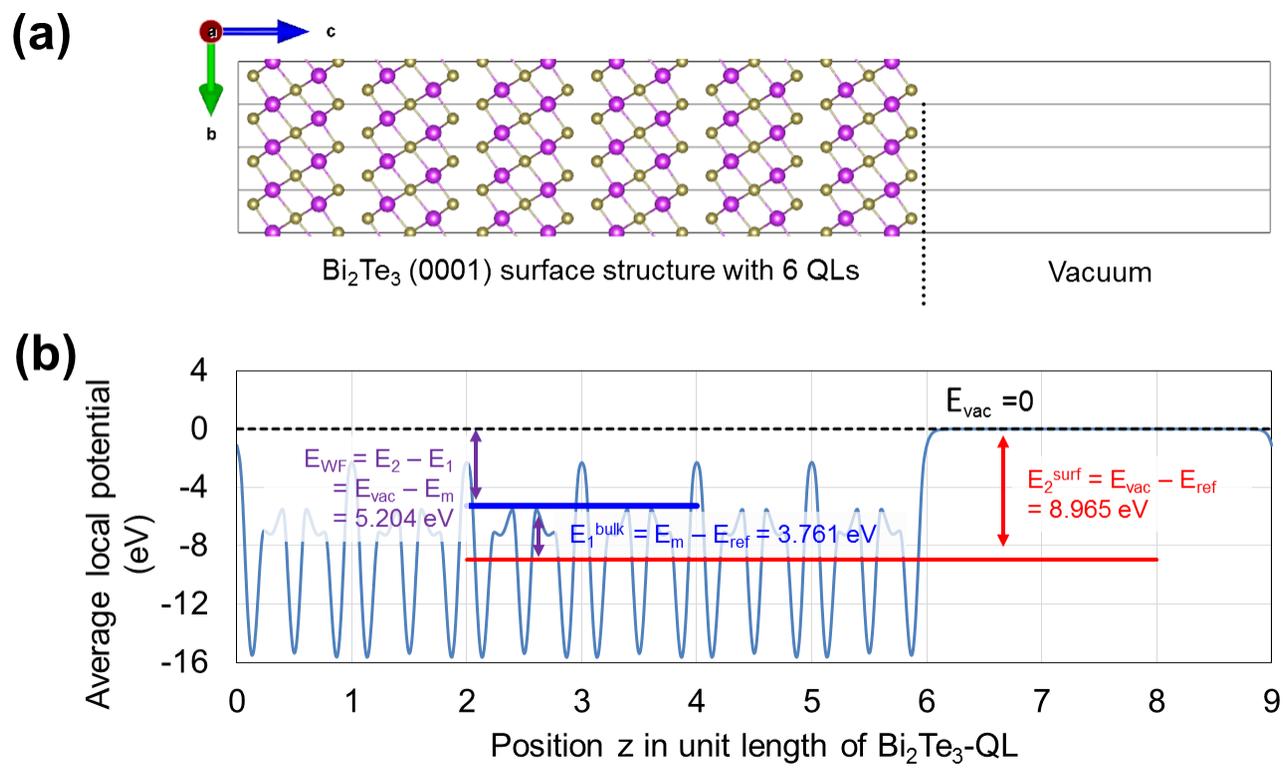

**Figure 3**

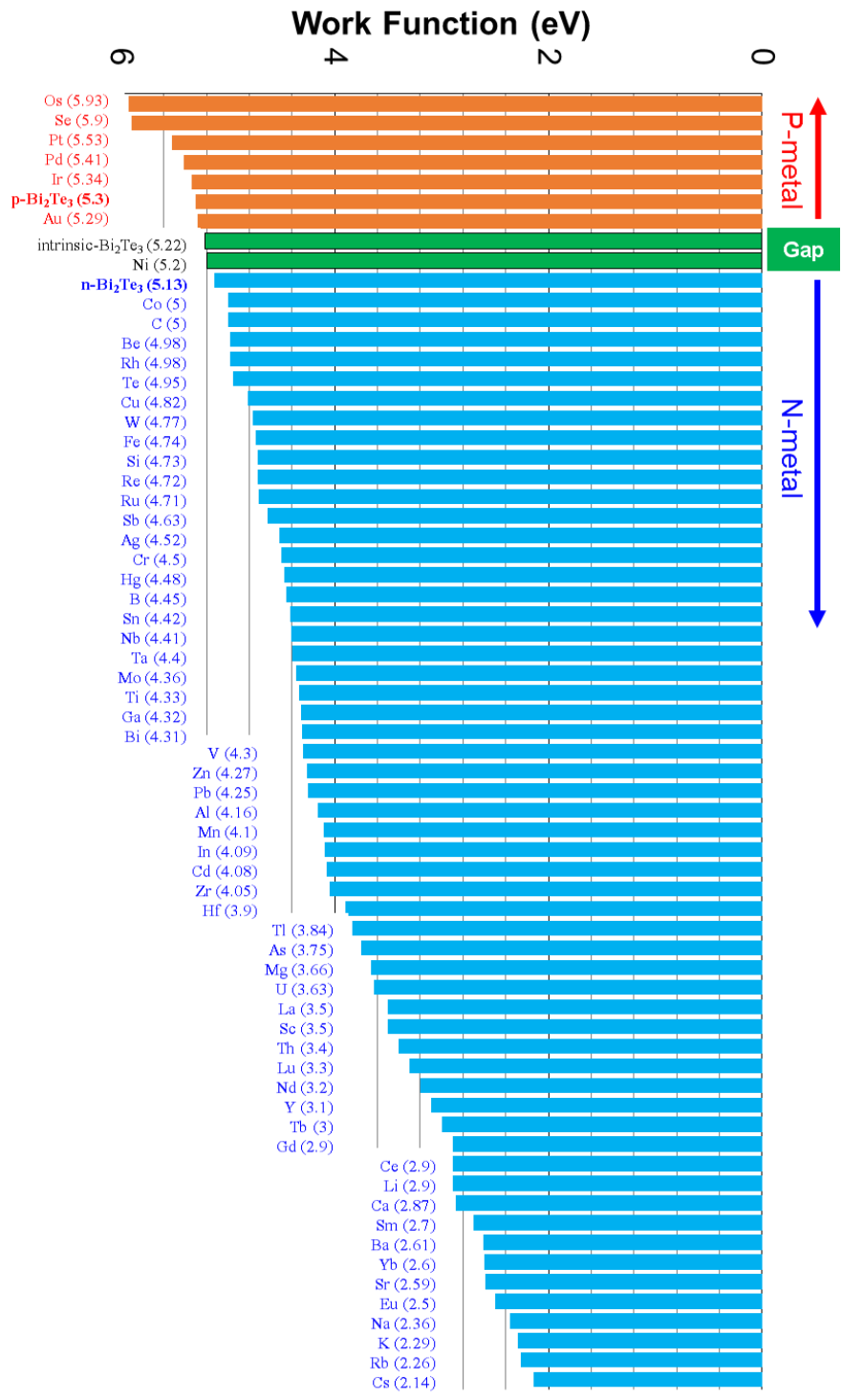

**Figure 4**